\documentclass[]{spie}  

\usepackage{amsmath,amsfonts,amssymb}
\usepackage{graphicx}
\usepackage[colorlinks=true, allcolors=blue]{hyperref}
\usepackage{soul}

\usepackage{caption} 
\newcommand{\fk}{f_k}

\title{Long-Timescale Stability in CMB Observations at Multiple Frequencies using Front-End Polarization Modulation}

\author[a]{Joseph Cleary}
\author[a]{Rahul Datta}
\author[a]{John W. Appel}
\author[a]{Charles L. Bennett}
\author[b]{David T. Chuss}
\author[a]{Jullianna Denes Couto}
\author[c]{Sumit Dahal}
\author[d]{Francisco Espinoza}
\author[c]{Thomas Essinger-Hileman}
\author[e]{Kathleen Harrington}
\author[a]{Jeffrey Iuliano}
\author[a]{Yunyang Li}
\author[a]{Tobias A. Marriage}
\author[a]{Carolina N\'{u}\~{n}ez}
\author[f]{Matthew A. Petroff}
\author[g]{Rodrigo A. Reeves}
\author[a]{Rui Shi}
\author[h]{Duncan J. Watts}
\author[c]{Edward J. Wollack}
\author[i]{Zhilei Xu}

\affil[a]{The William H. Miller III Department of Physics and Astronomy, Johns Hopkins University, 3701 San Martin Drive, Baltimore, MD
21218, USA}
\affil[b]{Department of Physics, Villanova University, 800 Lancaster Avenue, Villanova, PA 19085, USA}
\affil[c]{Goddard Space Flight Center, 8800 Greenbelt Road, Greenbelt, MD 20771, US}
\affil[d]{Departamento de Ingenier\'ia El\'ectrica, Universidad Cat\'olica de la Sant\'isima Concepci\'on, Alonso de Ribera 2850, Concepci\'on, Chile}
\affil[e]{Department of Astronomy and Astrophysics, University of Chicago, 5640 South Ellis Avenue, Chicago, IL 60637, USA}
\affil[f]{Center for Astrophysics, Harvard \& Smithsonian, 60 Garden Street, Cambridge, MA 02138, USA}
\affil[g]{CePIA, Astronomy Department, Universidad de Concepción, Casilla 160-C, Concepción, Chile}
\affil[h]{Institute of Theoretical Astrophysics, University of Oslo, P.O. Box 1029 Blindern, N-0315 Oslo, Norway}
\affil[i]{MIT Kavli Institute, Massachusetts Institute of Technology, 77 Massachusetts Avenue, Cambridge, MA 02139, USA}

\authorinfo{(Send correspondence to J.C.)\\E-mail: jcleary3@jhu.edu}

\begin{document} 
\maketitle

\begin{abstract}
The Cosmology Large Angular Scale Surveyor (CLASS) is a telescope array observing the Cosmic Microwave Background (CMB) at frequency bands centered near 40, 90, 150, and 220 GHz. CLASS measures the CMB polarization on the largest angular scales to constrain the inflationary tensor-to-scalar ratio and the optical depth due to reionization. To achieve the long time-scale stability necessary for this measurement from the ground, CLASS utilizes a front-end, variable-delay polarization modulator on each telescope. Here we report on the improvements in stability afforded by front-end modulation using data across all four CLASS frequencies. Across one month of modulated linear polarization data in 2021, CLASS achieved median knee frequencies of 9.1, 29.1, 20.4, and 36.4 mHz for the 40, 90, 150, and 220 GHz observing bands. The knee frequencies are approximately an order of magnitude lower than achieved via CLASS pair-differencing orthogonal detector pairs without modulation.
\end{abstract}


\section{INTRODUCTION}
\label{sec:intro}  

Originating only 380,000 years after the Big Bang, the Cosmic Microwave Background (CMB) is a direct window into the early Universe. The polarization of the CMB contains a wealth of information relating to inflation, reionization, dark matter interactions, and more.\cite{wmapResults,plank2018params,bicep,actBi} The Cosmology Large Angular Scale Surveyor (CLASS) measures the CMB polarization at frequency bands centered near 40, 90, 150, and 220 GHz.\cite{TEH2014,KH2016} Each CLASS telescope consists of a novel, front-end polarization modulator, custom optics and cryogenics, and a focal plane of transition-edge sensor (TES) bolometers\cite{jeff2018, dahal2022}. CLASS has observed from the Atacama Desert in Chile since 2016 with the goal of measuring the largest scales accessible from the ground. 

The first optical element in each CLASS telescope is a variable-delay polarization modulator (VPM), which consists of a polarizing wire grid in front of and parallel to a movable mirror.\cite{Chuss2012} This creates a phase delay between incident polarization parallel and perpendicular to the wires. The CLASS VPM mirrors move at a frequency of 10 Hz relative to the wire grid. This rapid modulation encodes the CMB linear polarization (Stokes $Q$/$U$) at a frequency above typical atmospheric and instrumental noise sources that would otherwise impact large angular-scale observations.\cite{miller2016} The VPM also provides sensitivity to circular polarization (Stokes $V$). The CLASS VPM design and 40 GHz performance have been described in Harrington et al. (2018)\cite{KH2018} and Harrington et al. (2021)\cite{KH2021}. In this proceeding we provide an analysis of an initial, representative data set at all four CLASS frequencies.

\section{Data \& Modeling}
\label{sec:data}

Most CMB polarimeters today utilize total power sensors (such as TESs) that are coupled (e.g., by probe antennas) to a single linear polarization. Thus these detectors measure a combination of the total intensity (Stokes $I$), linear polarization (Stokes $Q$/$U$), and circular polarization (Stokes $V$). Orthogonal detector pairs can be differenced to isolate the linear polarization signal. In principle, the pair-difference cancels the common, dominant unpolarized component and the circular polarization. In practice, the cancellation is imperfect and the pair-differenced data will also include spurious signal (i.e., intensity-to-polarization leakage) that could overwhelm the cosmological signal CLASS aims to measure. To suppress the spurious signal on the largest angular scales, CLASS modulates Stokes $U$ and $V$ at a frequency of 10 Hz. (Here, $+Q$ is defined parallel to the VPM wires.) The 10 Hz modulation frequency of the CLASS VPMs is much higher than the rate at which this spurious signal is expected to change, which can then be removed via high pass filtering. CLASS is then left with modulated $U$ and $V$. CLASS tracks the relative position of the VPM wire grid and mirror; thus, the modulation functions are known at all times. This allows the recovery of the incident $U$ and $V$ signals. To observe the $Q$ signal, CLASS performs daily boresight rotations that range between -45$^\circ$ and +45$^\circ$. For a more rigorous discussion of demodulation, see Harrington et al. (2021).\cite{KH2021} 

\begin{figure}[ht]
   \begin{center}
   \includegraphics[height=7cm]{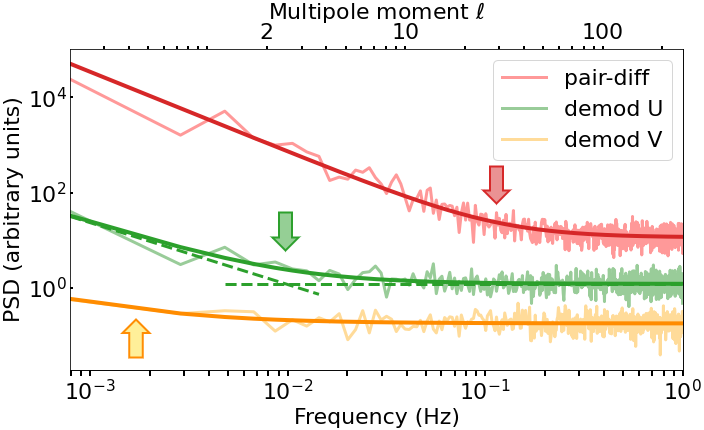}
   \end{center}
   \caption[example] 
   { \label{fig:fkExample} 
Spectra of 40 GHz data from one 3-hour segment, arbitrarily scaled and offset for visual clarity. Spectra are modeled as white noise plus a power law (dashed lines for demodulated-$U$). The knee frequency is where these components are equal (colored arrows). Demodulation significantly lowers the knee frequency compared to pair differencing alone, enabling access to the largest scales (lowest $\ell$, top axis).}
   \end{figure}

We perform a similar analysis to Harrington et al. (2021),\cite{KH2021} focusing on a representative subset of data collected during November 2021, when CLASS observed at all four frequencies. CLASS observes by performing continuous 720$^{\circ}$ rotations in azimuth at constant elevation. These observations were broken into 3-hour segments, then corrected for glitches and other errors. Data segments from known poorly performing detectors and high glitch rates are excluded. We also excluded segments when the average wind speed at the CLASS site was above 5 m/s and the precipitable water vapor (PWV) was above 3 mm.\footnote[2]{PWV data were acquired from the APEX radiometer website, \url{https://archive.eso.org/wdb/wdb/asm/meteo_apex/form}} After pair differencing and demodulation, power spectral densities are calculated for the pair-differenced, demodulated-$U$, and demodulated-$V$ data. Once spectra have been produced for each data segment, a noise model is fit to each spectrum. This model, given in Equation \ref{eq:noiseModel}, consists of two components: a constant white noise and a power law rising to low frequencies (red noise).
\begin{equation}
\label{eq:noiseModel}
PSD(f) = w_n^2 \Bigg(1 + \Bigg( \frac{f}{f_k} \Bigg)^\alpha \Bigg)
\end{equation}
The model parameters are the power law slope $\alpha$, the white noise level $w_n$, and knee frequency $\fk$, which is the frequency at which the two components of the model are equal. This is demonstrated in Figure \ref{fig:fkExample}, which shows spectra of CLASS pair-differenced, demodulated-$U$, and demodulated-$V$ data from a single 3-hour segment with arbitrary scale and offset for visual clarity. The dashed lines in the $U$ spectrum represent the separate red and white noise components. The colored arrows indicate the knee frequency of each spectra. Using a CLASS azimuthal scan speed of $\omega=2^{\circ}$/s and elevation $\theta=45^{\circ}$, each knee frequency $f$ can be converted into an angular scale on the sky and a corresponding multipole $\ell \approx 360^{\circ} f/ \omega\sin{\theta} $. The multipoles are shown in the top axis of Figure \ref{fig:fkExample}.

\section{Results}
\label{sec:results}

\begin{figure}[ht]
   \begin{center}
   \includegraphics[width=10cm]{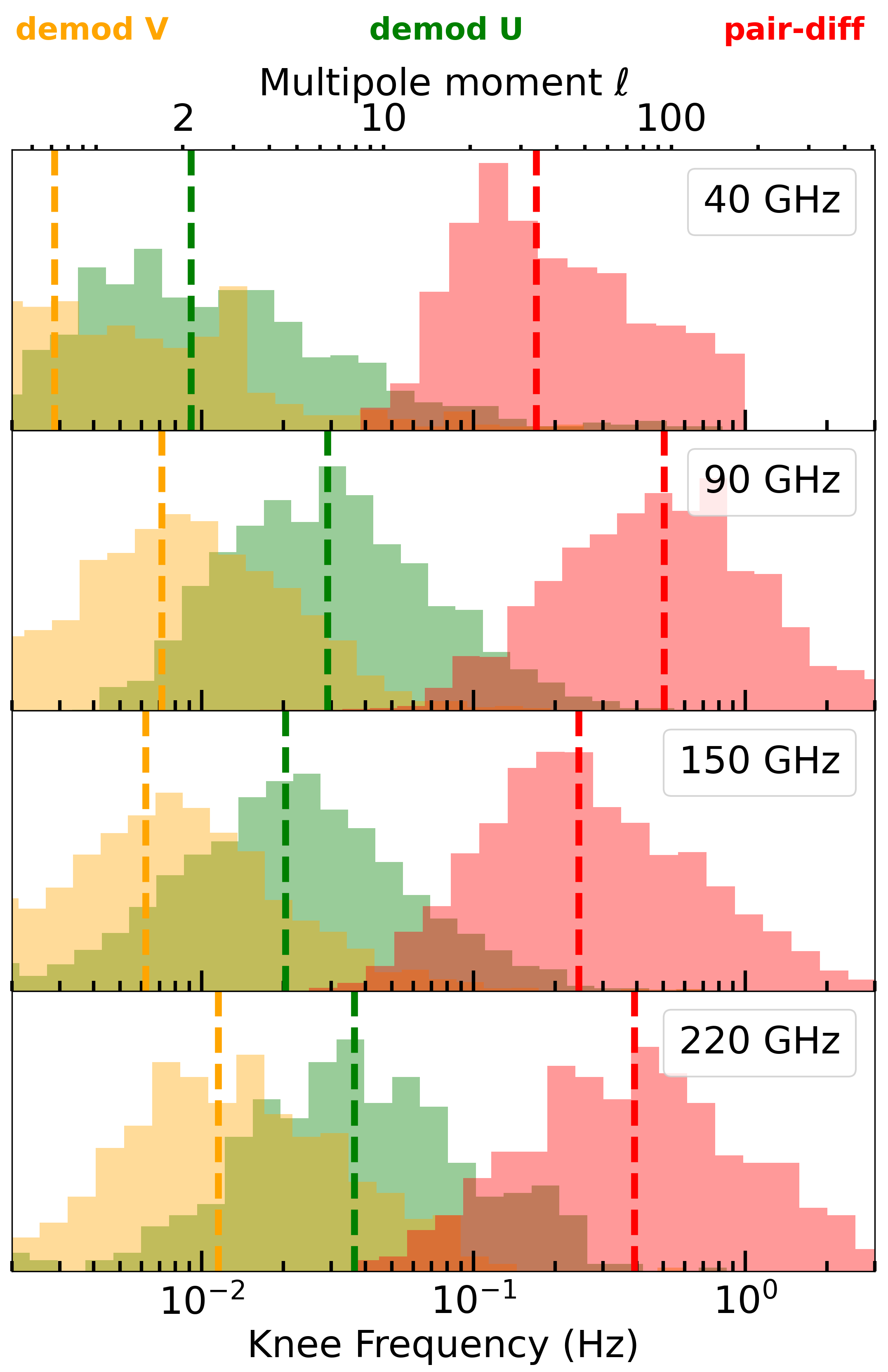}
   \end{center}
   \caption { \label{fig:dists} 
Distribution of knee frequencies for all four CLASS observing bands, for pair-differenced (red), demodulated-$U$ (green), and demodulated-$V$ (yellow). Medians of each distribution as denoted by vertical dashed lines.}
   \end{figure}

Data from all four observing bands were modeled to extract knee frequency values. Figure \ref{fig:dists} shows the distribution of knee frequencies for each band. The medians and 16$\%$ and 84$\%$ percentiles of each distribution are given in Table \ref{tab:fks}. For all four bands, the overall distribution of knee frequencies is significantly lower for demodulated compared to pair-differenced data. The median knee frequency decreases by approximately an order of magnitude between the pair-differenced and $U$ data. CLASS aims to constrain the CMB polarization down to multipoles $\ell < 10$. From Figures \ref{fig:fkExample} and \ref{fig:dists}, given the CLASS scan speed, this roughly corresponds to a knee frequency $\fk \lesssim 40\,\mathrm{mHz}$. The median knee frequency for all four CLASS bands is below 40 mHz for the $U$ data. The lower knee frequencies provided by demodulation significantly reduce the noise present in the CLASS data at low-$\ell$. The knee frequencies achieved by CLASS are necessary to measure the CMB on the largest angular scales. However, there are other systematic effects, such as scan-synchronous pick-up, present at low-$\ell$ that are the subject of separate studies.

\begin{table}[ht]
    \centering
    \begin{tabular}{|c|c|c|c|c|}
    \hline
    Band & \# segments & pair-diff (mHz) & demod-$U$ (mHz) & demod-$V$ (mHz)\\
    \hline
    40 GHz & 946 & 170$_{-78.9}^{+297}$ & 9.1$_{-5.4}^{+27}$ & 2.9$_{-2.0}^{+9.6}$ \\
    \hline
    90 GHz & 2429 & 503$_{-305}^{+752}$ & 29.1$_{-16.7}^{+41.9}$ & 7.1$_{-5.4}^{+12.1}$ \\
    \hline
    150 GHz & 3447 & 244$_{-134}^{+464}$ & 20.4$_{-13.3}^{+32.0}$ & 6.2$_{-4.7}^{+10.5}$ \\
    \hline
    220 GHz & 561 & 391$_{-243}^{+629}$ & 36.4$_{-22.1}^{+58.3}$ & 11.5$_{-7.1}^{+20.6}$ \\
    \hline
    \end{tabular}
    \caption{\label{tab:fks} Median knee frequencies for each CLASS observing band and data set from November 2021. Error values give the 68-percentile range of the knee frequency distribution around the median. The second column gives the number of 3-hour data segments analyzed for each band.}
\end{table}

As is clear from Figure \ref{fig:dists} and Table \ref{tab:fks}, the $V$ knee frequencies are systematically lower than those for $U$. This implies that, compared to $U$, the $V$ data has a higher white noise level relative to the red noise component. This may be due to higher $V$ white noise in the presence of comparable red noise, or due to overall weaker red noise in the CLASS $V$ data. Sources of noise in the polarization data that are not modulated by the VPM have their amplitudes increased by correcting the signal and data for the VPM modulation efficiency.\cite{KH2018} White noise for the $V$ data is approximately double the $U$ white noise due to this correction.\cite{KH2021} Whether the higher $V$ white noise is responsible for the lower knee frequencies depends on whether the red noise is uncorrelated with the $V$ modulation function or if it represents on-sky signal.\footnote[2]{While CLASS has made the first detection of continuum circular polarization due to Zeeman splitting of oxygen in the atmosphere\cite{petroff2020, padilla2020}, we do not expect this component to have significant random fluctuations, and there are no other known sources of $V$ fluctuations on sky.}

\section{Conclusions}
\label{sec:conc}

CLASS observes the CMB polarization on the largest angular scales to constrain many cosmological phenomena such as inflation and reionization. Employing a VPM on each telescope provides CLASS with additional observational stability beyond that of a standard, pair-differenced-only polarization measurement. This is demonstrated here by the reduction in knee frequencies of demodulated CLASS data compared to pair differencing alone. For the data considered here, CLASS achieved median knee frequencies of 9.1, 29.1, 20.4, and 36.4 mHz for the $U$ data at 40, 90, 150, and 220 GHz, respectively. While red noise is not the sole factor in the final low-$\ell$ sensitivity of CLASS, these results provide a first look at the long-timescale stability of CLASS data across all four frequencies. On-going work will expand this analysis to include the whole multi-frequency data set.

\acknowledgments
We acknowledge the National Science Foundation Division of Astronomical Sciences for their support of CLASS under Grant Numbers 0959349, 1429236, 1636634, 1654494, 2034400, and 2109311. The CLASS project employs detector technology developed under several previous and ongoing NASA grants. Detector development work at JHU was funded by NASA cooperative agreement 80NSSC19M0005. Data analysis for CLASS is conducted using computational resources of the Advanced Research Computing at Hopkins (ARCH). We further acknowledge the very generous support of Jim and Heather Murren (JHU A$\&$S ’88), Matthew Polk (JHU A$\&$S Physics BS ’71), David Nicholson, and Michael Bloomberg (JHU Engineering ’64). R.R. is supported by the ANID BASAL projects ACE210002 and FB210003. Z.X. is supported by the Gordon and Betty Moore Foundation through grant GBMF5215 to the Massachusetts Institute of Technology.  CLASS is located in the Parque Astronómico Atacama in northern
Chile under the auspices of the Agencia Nacional de Investigación y Desarrollo (ANID).

\bibliography{report} 
\bibliographystyle{spiebib} 

\end{document}